\documentclass[pdflatex,sn-mathphys-num]{sn-jnl}

\usepackage{graphicx}%
\usepackage{multirow}%
\usepackage{amsmath,amssymb,amsfonts}%
\usepackage{amsthm}%
\usepackage{mathrsfs}%
\usepackage[title]{appendix}%
\usepackage{xcolor}%
\usepackage{textcomp}%
\usepackage{manyfoot}%
\usepackage{booktabs}%
\usepackage{listings}%
\usepackage{braket}
\usepackage{array}
\usepackage{makecell}
\usepackage{caption}
\usepackage{subcaption}
\usepackage{hyperref}

\usepackage[ruled,vlined,linesnumbered]{algorithm2e}
\SetKwInOut{Input}{Input}
\SetKwInOut{Output}{Output}

\theoremstyle{thmstyleone}%
\theoremstyle{thmstyletwo}%

\theoremstyle{thmstylethree}%

\begin{document}

\title{Structural encoding with classical codes for computational-basis bit-flip correction in the early fault-tolerant regime}

\author*[1]{\fnm{IlKwon} \sur{Sohn}}\email{d2estiny@kisti.re.kr}

\author[1]{\fnm{Changyeol} \sur{Lee}}\email{lcy253898@kisti.re.kr}

\author[1]{\fnm{Wooyeong} \sur{Song}}\email{wysong@kisti.re.kr}

\author[1]{\fnm{Kwangil} \sur{Bae}}\email{kibae@kisti.re.kr}

\author[1]{\fnm{Wonhyuk} \sur{Lee}}\email{livezone@kisti.re.kr}

\affil[1]{\orgdiv{Quantum Network Research center}, \orgname{Korea Institute of Science and Technology Information}, \orgaddress{\city{Daejeon}, \postcode{34141},\country{Republic of Korea}}}

\abstract{
Achieving reliable performance on early fault-tolerant quantum hardware will depend on protocols that manage noise without incurring prohibitive overhead.
We propose a novel framework that integrates quantum computation with the functionality of classical error correction.
In this approach, quantum computation is performed within the codeword subspace defined by a classical error correction code.
The correction of various types of errors that manifest as bit flips is carried out based on the final measurement outcomes.
The approach leverages the asymmetric structure of many key algorithms, where problem-defining diagonal operators (e.g., oracles) are paired with fixed non-diagonal operators (e.g., diffusion operators).
The proposed encoding maps computational basis states to classical codewords.
This approach commutes with diagonal operators, obviating their overhead and confining the main computational cost to simpler non-diagonal components.
Noisy simulations corroborate this analysis, demonstrating that the proposed scheme serves as a viable protocol-level layer for enhancing performance in the early fault-tolerant regime.
}

\maketitle

\section*{Introduction}
Quantum computing has demonstrated the potential to offer computational advantages over classical computation for certain problems~\cite{de92, sh94, si97, ha09}.
However, the performance of quantum algorithms on current quantum devices is severely constrained by hardware-level noise~\cite{pr18, bh22}.
While quantum error correction (QEC) provides a theoretical framework for suppressing such errors, its practical implementation requires high-fidelity physical qubits, numerous ancilla qubits, and complex gate operations, which are not yet broadly available on near-term devices~\cite{sh95, kn05, fo012, te15, ca17, ch18}.

To overcome these limits, various quantum error mitigation schemes have been proposed to reduce the impact of noise on such hardware~\cite{te17, li17, ca23}.
Zero Noise Extrapolation (ZNE)~\cite{te17, li17, gi20}, Probabilistic Error Cancellation (PEC)~\cite{te17, va23}, Virtual Distillation (VD)~\cite{hu21, ko21}, Measurement Error Mitigation (MEM)~\cite{na20, br21, gu21, hi22}, and Symmetry Verification (SV)~\cite{bo18, sa19, ca21, zh24} are representative examples.
However, most of these approaches rely on statistical post-processing of full output distributions or require precise prior knowledge of the underlying noise model~\cite{te17, na20, ca23}.
These features make them less effective when only a limited number of measurement shots are available, and they often involve computationally intensive procedures that are unsuitable for real-time applications~\cite{ta22, ca23}.

This work introduces a framework for enhancing the performance of quantum algorithms by integrating the structure of classical error correction codes (ECCs) directly into the computation.
The proposed scheme mitigates the effects of various quantum errors that occur during computation by correcting bit-flip errors in the final measurement outcomes using classical decoding.
The core idea is to perform the computation within the codeword subspace defined by a systematic classical linear code.
This approach is particularly effective for algorithms with an asymmetric operational structure, such as Grover's search algorithm~\cite{gr96, gr97} and the Quantum Approximate Optimization Algorithm (QAOA)~\cite{fa14, ha19}.
In such algorithms, the encoding commutes with the complex, problem-defining diagonal operators (e.g., oracles), incurring zero overhead for their execution, while the main implementation overhead is strategically confined to the simpler, fixed non-diagonal operators (e.g., diffusion operators).

This method offers practical advantages over existing mitigation schemes.
Unlike methods that rely on statistical post-processing like ZNE and PEC, the scheme's per-shot classical decoding is effective even with limited measurement shots and suitable for hybrid workflows requiring real-time feedback.
Furthermore, it provides more comprehensive protection than recent  MEM schemes that only address the final readout process~\cite{gu21, hi22}, by providing resilience against errors that accumulate during the execution of the most critical operators.
The proposed scheme does not require resource-intensive components of QEC, such as logical ancilla qubits or syndrome measurements.
It is particularly well-suited for architectures with all-to-all connectivity, such as trapped ions~\cite{br19, wr19, he20, bl21, ch24} and neutral atoms~\cite{eb22, ev23, bl24}, where the encoding overhead is minimized.
These characteristics establish the protocol as a candidate for a complementary layer in the early fault-tolerant regime, which will require practical methods for suppressing residual errors.
We demonstrate through noisy simulations that this scheme is particularly effective for mitigating errors in exemplary cases of medium-scale circuits, where the relative cost of encoding becomes negligible while the protective benefits are amplified.

\section*{Results}
\label{res}
\subsection*{Protocol description}
The proposed scheme operates within a quantum computational framework, but is built upon the principles of classical error correction codes (ECCs).
This intersection requires a careful choice of terminology to avoid ambiguity with the language of fault-tolerant quantum computing (FTQC), such as the term \textit{logical}.

To maintain clarity, we will adopt terminology inspired by classical ECC theory.
In classical coding, the original data is called a message, and the encoded data is a codeword.
While the proposed scheme directly uses the concept of a codeword, the term `message' is more suited to a communication context.
As our focus is on computations, we will use a more fitting term for the unencoded data.
Therefore, for the remainder of this paper, we establish the following terminology:
\begin{itemize}
    \item The algorithm's unencoded $k$-qubit space is termed the \textbf{computational space}, and a state within it is a \textbf{computational state}.
    \item The $n$-qubit space spanned by the classical codewords is termed the \textbf{codeword space}, and a state within it is a \textbf{codeword state}.
\end{itemize}

\subsubsection*{Systematic ECC encoding}
The protocol begins by defining an isometric mapping from the $k$-qubit computational Hilbert space, $\mathcal{H}_k$, to the codeword subspace within the $n$-qubit Hilbert space, $\mathcal{H}_n$.
This mapping is based on a classical linear systematic $[n, k, d]$ code over $\mathbb{F}_2$, where $d$ is the minimum distance that determines the code's error-correction capability.
The code is characterized by a generator matrix $G = [I_k | P]$, where $I_k$ is the $k \times k$ identity matrix and $P\in \{0,1\}^{k\times(n-k)}$ \cite{li83}.

A computational basis state $\ket{x}$, where $x \in \{0,1\}^k$, is mapped to a codeword state $\ket{C(x)}$, which corresponds to a classical codeword.
A key property of a systematic code is that the codeword's structure is partitioned.
The first $k$ qubits preserve the original computational bits, while the remaining $n-k$ qubits encode the parity bits, $p(x) = xP$, computed over $\mathbb{F}_2$.
Consequently, the codeword state has the tensor product structure $\ket{C(x)} = \ket{x} \otimes \ket{p(x)}$.

This encoding is implemented by a unitary operator, $U_E$, which prepares the codeword state such that
\begin{equation}
    U_E (\ket{x} \otimes \ket{0}^{n-k}) = \ket{x} \otimes \ket{p(x)} = \ket{C(x)}.
\end{equation}
This unitary transformation effectively embeds the $k$-qubit computational space into the $n$-qubit codeword space, establishing the foundation for the error mitigation protocol.

\subsubsection*{Asymmetric implementation of operators}
The proposed scheme minimizes its encoding overhead through an asymmetric implementation of operators.
This asymmetric implementation allows complex diagonal operators to be applied with zero overhead and to directly benefit from the error-resilience of the codeword subspace, while confining the main computational overhead to the simpler non-diagonal operators.

\paragraph*{Diagonal operators}
A diagonal operator $O_D$ on the computational space applies a phase based on the basis state: $O_D\ket{x} = e^{i\phi_x}\ket{x}$.
This operator can be applied to a codeword state with zero overhead.
The corresponding operator in the codeword subspace, $O_D^E$, is the original computational operator acting on the first $k$ qubits and the identity on the remaining $n-k$ qubits:
\begin{equation}
    O_D^E = O_D \otimes I_{n-k}.
\end{equation}
When applied to a codeword state, the effect is equivalent to the computational operator:
\begin{equation}
    O_D^E \ket{C(x)} = (O_D \otimes I_{n-k}) \ket{x} \otimes \ket{p(x)} = e^{i\phi_x}\ket{C(x)}.
\end{equation}
Because the encoding operator $U_E$ commutes with the diagonal operator $O_D^E$, there is no need for constructing an encoded operator.
Consequently, the overhead associated with implementing diagonal operators is obviated.

\paragraph*{Non-diagonal operators}
A non-diagonal operator $O_{ND}$ must be conjugated by the encoding unitary $U_E$ to be correctly applied within the codeword subspace.
The corresponding operator $O_{ND}^E$ is:
\begin{equation}
    O_{ND}^E = U_E (O_{ND} \otimes I_{n-k}) U_E^\dagger.
\end{equation}
This conjugation, which decodes, operates, and re-encodes the state, constitutes the main implementation overhead of the protocol.

\begin{figure}[tb!]
\centerline{\includegraphics[width=0.8\linewidth]{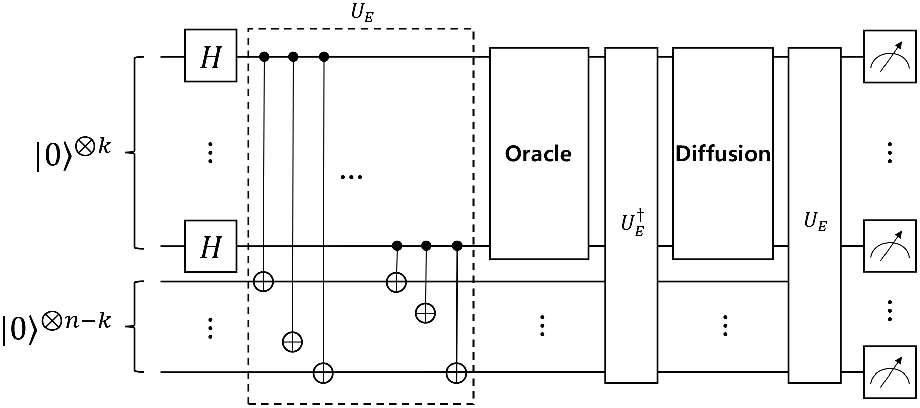}}
    \caption{A conceptual circuit diagram for Grover's search algorithm using the proposed scheme.
    The system employs $k$ computational and $n-k$ ancilla qubits.
    The core feature is the asymmetric implementation: the diagonal oracle is applied directly with zero overhead, while the non-diagonal diffusion is implemented by conjugation with $U_E$ and $U_E^\dagger$.
    Final measurement outcomes are classically decoded to correct computational errors.}
    \label{fig:proex}
\end{figure}

\subsubsection*{Measurement and classical decoding}
Upon completion of the quantum computation, a measurement is performed on the $n$ qubits.
The classical decoding algorithm then attempts to map each measured bitstring back to a valid codeword.
If an outcome has been affected by errors beyond the code's correction capability, it will fail to decode and is subsequently discarded.

The efficacy of this decoding process extends beyond the guaranteed error-correction capability of $t = \lfloor(d-1)/2\rfloor$~\cite{li83}.
The classical decoder can successfully correct an error pattern with a weight $w > t$, provided that the resulting syndrome is unique and not degenerate with the syndrome of any lower-weight error.
Increasing the code length $n$ for a fixed minimum distance $d$ expands the syndrome space by adding more parity-check constraints.
This larger, more structured space can increase the likelihood that such higher-weight, uniquely correctable errors exist, thereby enhancing the overall decoding performance.
This potential for improved correction must, however, be weighed against the increased physical error rates associated with the larger quantum circuits required for longer codes.
This trade-off implies that for any given hardware noise model, there exists an optimal code length $n$ that maximizes the scheme's performance.

This post-selection mechanism is conceptually analogous to Symmetric Verification~\cite{bo18, sa19, ca21, zh24}, another error mitigation scheme that discards measurement outcomes violating a known physical symmetry.
This decoding process allows the proposed scheme to mitigate errors on a per-shot basis, without statistical post-processing.
This entire process, from the initial encoding to the final decoding, is conceptually illustrated for Grover's search algorithm in Fig.~\ref{fig:proex}.

\subsection*{Simulation Results}
\subsubsection*{Application to Grover's search algorithm}
To comprehensively evaluate the proposed scheme, we designed three simulation scenarios focusing on Grover's search algorithm.
This algorithm serves as an ideal benchmark as it exemplifies the asymmetric operational structure that the proposed scheme targets: a problem-defining, diagonal oracle paired with a fixed, problem-independent diffusion operator.
The scenarios are as follows: First, we assess the scheme's robustness by varying physical error rates.
Second, we investigate the direct impact of the code's error-correction capability by comparing codes with different minimum distances.
Finally, we examine the scheme's scalability by applying it to problems of increasing computational size $k$.

To quantify the performance in these scenarios, we evaluate the scheme's two-fold enhancement mechanism: direct error correction and post-selection.
Error correction increases the number of correct outcomes, while post-selection acts as a filter, discarding corrupted shots.
Therefore, for evaluating Grover's search algorithm, we will report three key metrics: the conditional success probability ($P_{s|acc}$), defined as the success rate among accepted shots; the unconditional success probability ($P_s$), the success rate across all initial shots; and the acceptance rate ($A$ in \%), the fraction of shots that are not discarded.

By convention, the unmitigated baseline accepts all shots ($A=100\%$), hence $P_{s|acc}^{\mathrm{base}}=P_s^{\mathrm{base}}$.
For the proposed scheme, post-selection yields $0\le A\le 100$, so we report both $P_{s|acc}^{\mathrm{mit}}$ and $P_s^{\mathrm{mit}}$.
These quantities are related by the following identity.
\begin{equation}
    P_s^{\mathrm{mit}} = \frac{A}{100} \times P_{s|acc}^{\mathrm{mit}}.
    \label{eq:ps_relation}
\end{equation}
Unless otherwise stated, success probability refers to the conditional metric $P_{s|acc}$, whereas we explicitly write unconditional success probability when referring to $P_s$.
All differences are expressed in percentage points (pp), and acceptance is reported in percent (\%).

\paragraph*{Performance under varying noise levels}
\begin{figure}[t!]
    \centering
    \includegraphics[width=0.9\linewidth]{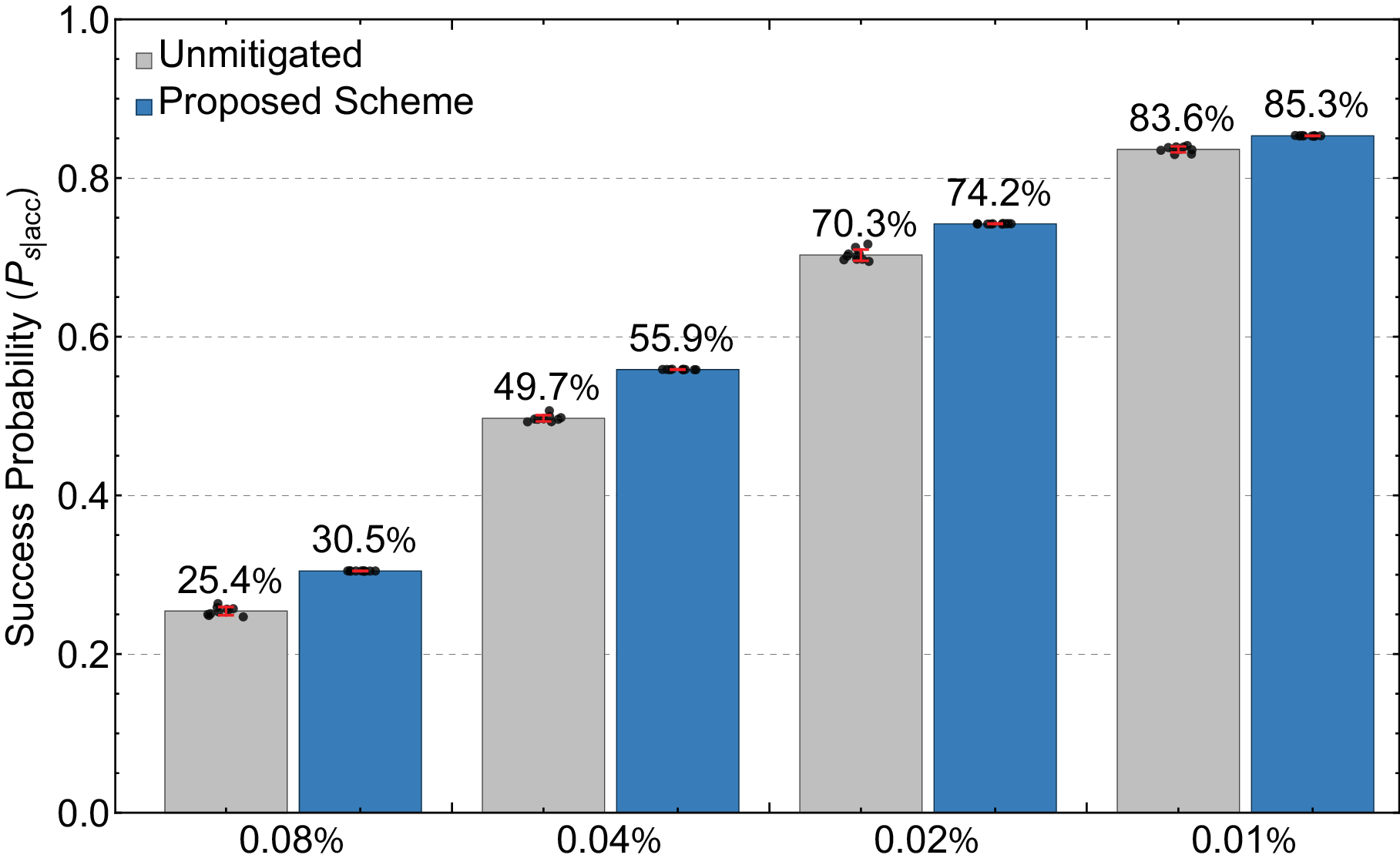}
    \caption{Performance of the proposed scheme using the systematic $[13,7,3]$ code under varying two-qubit gate error rates for a 7-qubit Grover's search algorithm.
    Results are aggregated over 10 random seeds with 4{,}000 shots per seed (40{,}000 shots per configuration).
    The proposed scheme (blue) consistently boosts the mean conditional success probability ($P_{s|acc}^{\mathrm{mit}}$) compared to the unmitigated baseline ($P_{s|acc}^{\mathrm{base}}$), with absolute gains ranging from +1.7 to +6.2 percentage points.
    Error bars represent one standard deviation over the 10 random seeds.}
    \label{fig:perf_n}
\end{figure}

We first benchmark the 7-qubit Grover's search algorithm targeting $\ket{1111111}$ at a noise level representative of current ion-trap systems: a single-qubit depolarizing error of $0.02\%$, a two-qubit error of $0.4\%$, and a readout error of $0.5\%$~\cite{ch24, IQFPage}.
Under these conditions, the unmitigated algorithm achieves an unconditional success probability $P_{s}^{\mathrm{base}}$ of only 0.87\%.
This is negligibly better than the random selection rate of 0.781\%.
This result highlights that for current hardware, the algorithm operates in a noise-dominated regime.
This motivates our main analysis, for which we explore improved physical fidelities over a range of two-qubit error rates from $0.08\%$ down to $0.01\%$.

The combined effect of error correction and post-selection leads to a consistent improvement in success probability across all tested noise levels, as shown in Fig.~\ref{fig:perf_n}.
For instance, at a 0.04\% error rate, the proposed scheme produces on average 97.0 more correct shots while operating at an acceptance rate of $A=93.3\%$ (see Table~\ref{tab:sim_n}).
This performance gain is achieved with a modest increase in circuit overhead; the two-qubit gate depth, after transpilation (as defined in the~\nameref{sys} section), increases from 4498 for the baseline circuit to 4565 for the circuit implementing the proposed scheme.
This increase corresponds to a relative overhead of approximately $1.5\%$, confirming that the benefits of error mitigation outweigh the implementation cost, even for medium-scale circuits such as the 7-qubit Grover’s search algorithm.
At a $0.08\%$ two-qubit error rate, the conditional success probability increases from $P_{s|acc}^{\mathrm{base}}=25.4\%$ to $P_{s|acc}^{\mathrm{mit}}=30.5\%$ (+5.1 pp).
At $0.04\%$, it increases from $P_{s|acc}^{\mathrm{base}}=49.7\%$ to $P_{s|acc}^{\mathrm{mit}}=55.9\%$ (+6.2 pp).
As the physical error rate decreases, the filtering aspect of the scheme becomes less critical.
This is reflected in the acceptance rate $A$ in Table~\ref{tab:sim_n}, which rises from $85.2\%$ to $98.3\%$ as the error rate drops from $0.08\%$ to $0.01\%$.
Consequently, while the scheme maintains its advantage, the relative gain in success probability narrows in the low-noise regime.

\begin{table}[h!]
\centering
    \caption{Performance comparison and circuit overhead for the 7-qubit Grover's search algorithm under various two-qubit gate error rates, using the systematic $[13,7,3]$ classical error correction code.
    The table includes the two-qubit gate depth to quantify the implementation overhead.
    Values are averaged over 10 runs with 4,000 shots, showing how the proposed scheme achieves a higher correct shot count and improved success probability through per-shot error correction and post-selection.}
    \label{tab:sim_n}
    \begin{tabular}{ccccccc}
    \toprule
    \multirow{2}{*}[-2ex]{\makecell{Two-qubit \\ error rate}} &
    \multicolumn{2}{c}{Unmitigated} &
    \multicolumn{4}{c}{Proposed scheme ($[13,7,3]$)} \\
    \cmidrule(lr){2-3} \cmidrule(lr){4-7}
    & \makecell{2Q gate \\ depth} & \makecell{Avg. correct \\ shots} & \makecell{2Q gate \\ depth} & \makecell{Avg. correct \\ shots} & \makecell{Avg. valid \\ shots} & \makecell{Acceptance \\ (\%)} \\
    \midrule
    0.08\% & \multirow{4}{*}{4498} & 1017.0 & \multirow{4}{*}{4565} & 1038.9 & 3408.9 & 85.2 \\
    0.04\% & & 1988.7 & & 2085.7 & 3733.0 & 93.3 \\
    0.02\% & & 2811.8 & & 2858.6 & 3850.8 & 96.3 \\
    0.01\% & & 3344.5 & & 3354.7 & 3931.8 & 98.3 \\
    \bottomrule
    \end{tabular}
\end{table}

\paragraph{Impact of encoding overhead on performance}
The choice of an error correction code presents a critical trade-off between its theoretical error-correction capability and the practical gate overhead required for its implementation.
A code with a larger minimum distance ($d$) can correct more errors but demands a more complex encoding circuit, which in turn introduces more noise.
This section investigates this trade-off for a 7-qubit computational space ($k=7$) to determine the optimal balance between protection and overhead.
\begin{figure}[t!]
    \centering
    \includegraphics[width=0.85\linewidth]{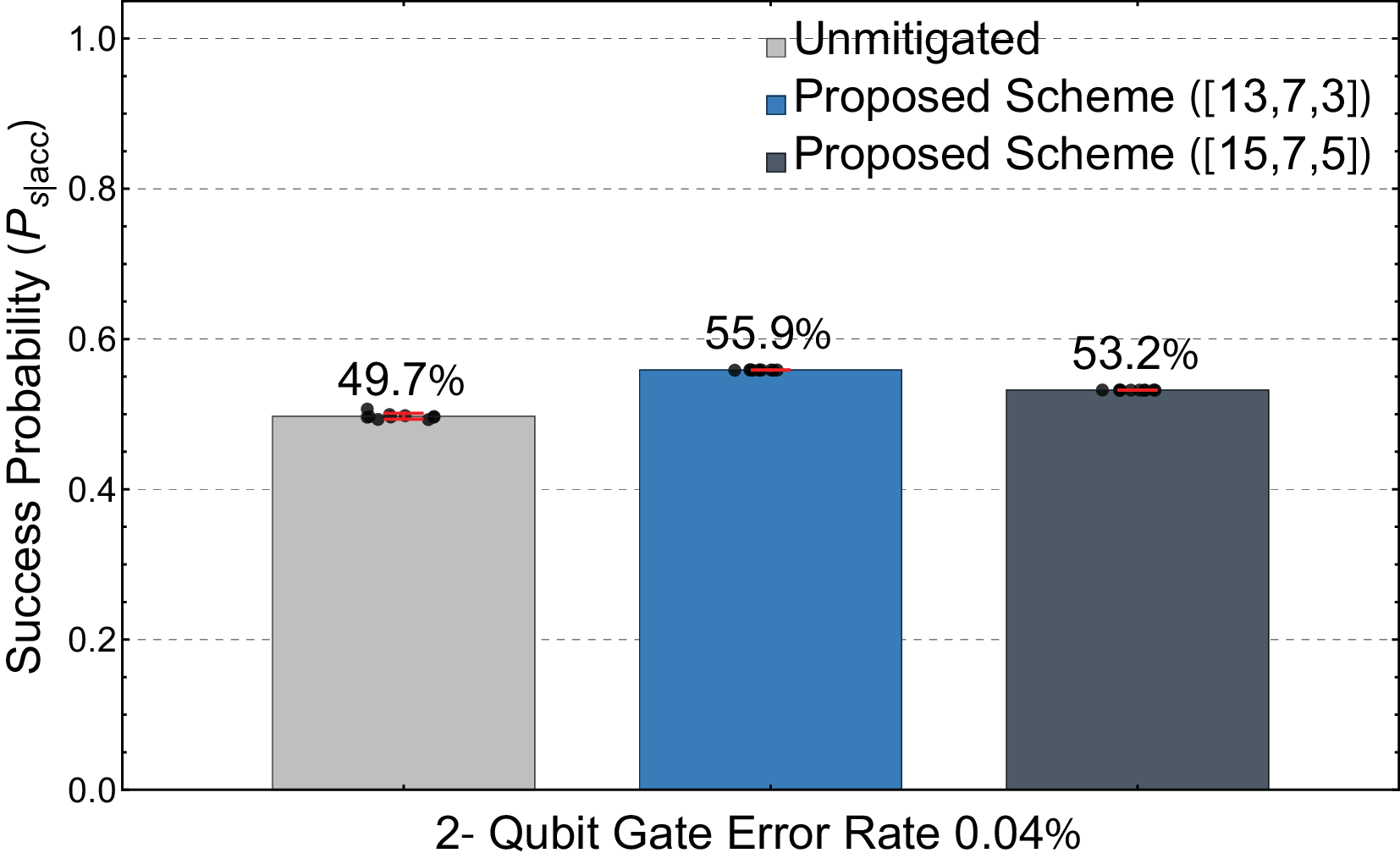}
    \caption{Impact of code strength and overhead for a 7-qubit problem ($k=7$) at a two-qubit gate error rate of $0.04\%$.
    The figure compares the conditional success probability ($P_{s|acc}^{\mathrm{mit}}$) of the lower-overhead $[13,7,3]$ ($d=3$) code with the stronger $[15,7,5]$ ($d=5$) code.
    Results are aggregated over 10 random seeds (40{,}000 total shots), and error bars represent one standard deviation.
    The superior performance of the $d=3$ code indicates that the gate overhead from the more complex encoding outweighs the benefit of a larger code distance in this regime.}
    \label{fig:perf_d}
\end{figure}

The unmitigated baseline attains a conditional success probability $P_{s|acc}^{\mathrm{base}}=49.7\%$.
With the proposed scheme using the systematic $[13,7,3]$ code ($d=3$), this rises to $P_{s|acc}^{\mathrm{mit}}=55.9\%$.
However, using the stronger $[15,7,5]$ code ($d=5$) yields a lower success probability of $P_{s|acc}^{\mathrm{mit}}=53.2\%$, as shown in Fig.~\ref{fig:perf_d}.
This lower performance is explained by the considerable gate overhead incurred by the more complex encoding circuit.
The encoding for the $[15,7,5]$ code requires 30 $\mathrm{CNOT}$ gates, more than double the 14 $\mathrm{CNOT}$s needed for the $[13,7,3]$ code.
For the full 7-qubit Grover's search algorithm, which consists of an initial encoding and eight Grover iterations, this difference is amplified.
The $d=3$ code adds a total of 238 additional $\mathrm{CNOT}$ gates to the baseline circuit, whereas the $d=5$ code adds 510.
In the tested noise regime, the accumulated errors from the extra 272 $\mathrm{CNOT}$s negate the benefits of its stronger error correction capability.
This finding demonstrates that a moderately strong code with lower overhead can be a more practical and effective choice.

\paragraph{Impact of the parity-bit budget on mitigation performance}
\begin{figure}[t!]
    \centering
    \includegraphics[width=0.9\linewidth]{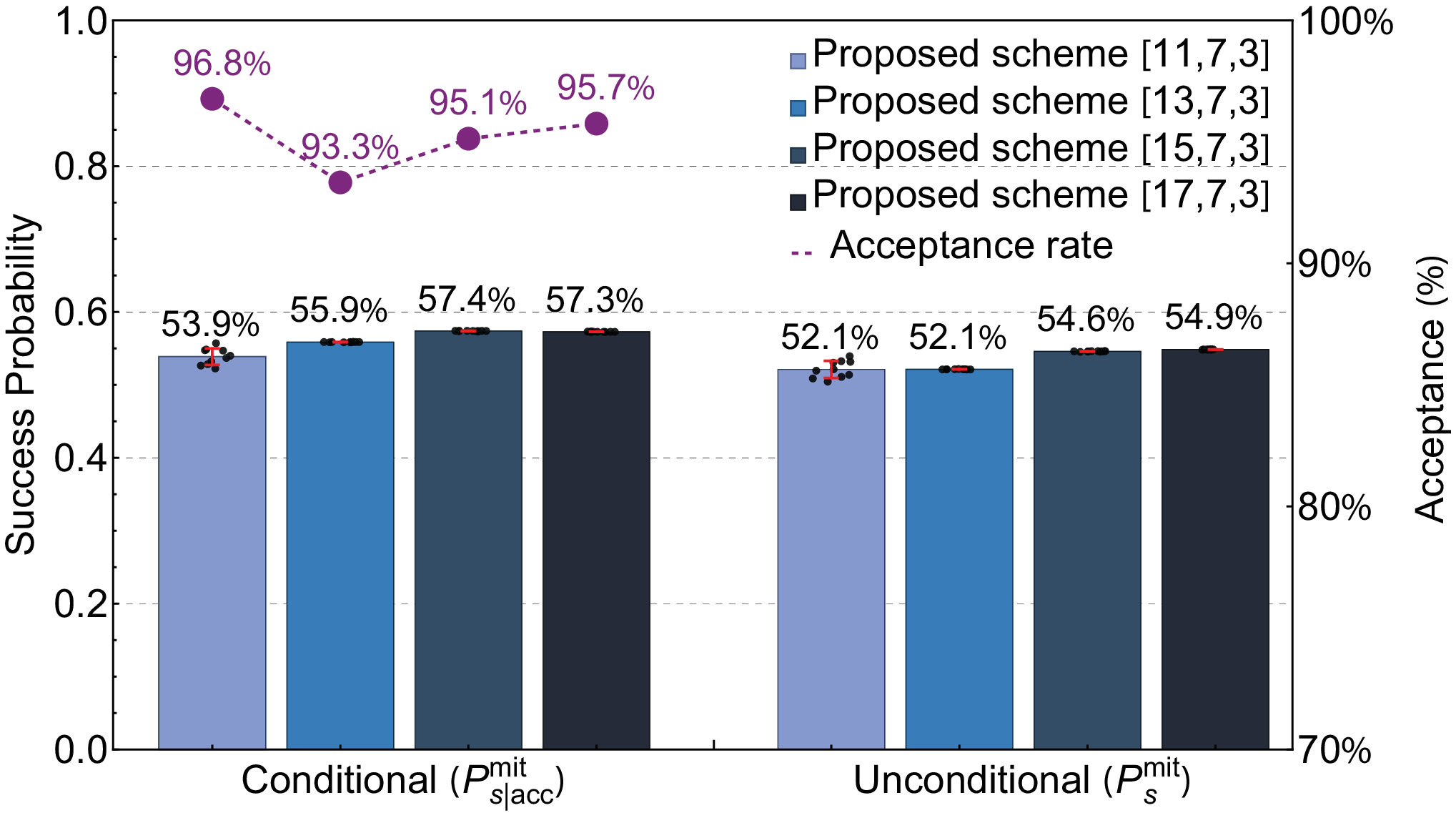}
    \caption{Impact of the parity-bit budget ($n-k$) on mitigation performance for a 7-qubit Grover's search algorithm ($d=3$) at a two-qubit gate error rate of 0.04\%.
    Results are aggregated over 10 random seeds with 4{,}000 shots per seed (40{,}000 shots per configuration).
    The figure reports the conditional ($P_{s|acc}^{\mathrm{mit}}$) and unconditional ($P_{s}^{\mathrm{mit}}$) success probabilities, with the Acceptance (\%) shown on the right $y$-axis.
    Error bars represent one standard deviation over the 10 random seeds.
    The $[17,7,3]$ code attains the highest unconditional performance ($54.9\%$), while $[15,7,3]$ offers the best practical trade-off between performance and circuit overhead.}
    \label{fig:perf_nk}
\end{figure}
Proceeding from the observation that a minimum distance of $d=3$ offers a practical balance between corrective capability and implementation overhead, this section investigates the impact of the parity-bit budget.
The analysis examines the performance variation resulting from an increasing number of parity bits ($n-k$) by changing the code length $n$, while maintaining a fixed computational space ($k=7$) and minimum distance ($d=3$).
As illustrated in Fig.~\ref{fig:perf_nk}, we evaluate performance with two metrics: the conditional success probability $P_{s|acc}$, computed from post-selected outcomes, and the unconditional success probability $P_{s}$, computed over the total set of initial shots.
The latter serves as a more holistic measure of the scheme's end-to-end performance.

The results reveal a nuanced trade-off between the two metrics.
The conditional metric indicates that the $[15,7,3]$ and $[17,7,3]$ codes achieve the highest values, with $P_{s|acc}^{\mathrm{mit}}=57.4\%$ and $P_{s|acc}^{\mathrm{mit}}=57.3\%$, respectively.
However, this metric does not fully capture the associated sampling cost.
This is clearly demonstrated by the $[13,7,3]$ code, which has the lowest Acceptance of 93.3\% among the tested codes.
The aggressive post-selection (smaller denominator) contributes to its elevated conditional success rate, but at the cost of discarding more shots.
A more comprehensive assessment is provided by the unconditional success probability (right side of Fig.~\ref{fig:perf_nk}), which shows that performance increases monotonically with $n$, culminating in $P_{s}^{\mathrm{mit}}=54.9\%$ for the $[17,7,3]$ code, slightly surpassing $P_{s}^{\mathrm{mit}}=54.6\%$ for the $[15,7,3]$ code.

The trend in unconditional probability reveals an optimal parity-bit budget based on overhead consideration.
Although the $[17,7,3]$ code provides the highest unconditional success probability (54.9\%), the minimal 0.3 pp gain over the $[15,7,3]$ code (54.6\%) does not justify its larger circuit overhead.
The $[15,7,3]$ code therefore represents the most practical optimum, balancing its strong error correction against its lower implementation cost compared to the $[17,7,3]$ code.
This underscores that the optimal strategy involves not a simple maximization of redundancy, but a careful budgeting of parity bits to maximize the unconditional probability of success relative to the circuit complexity.

\subsubsection*{Application to IQP sampling circuits}
To demonstrate the versatility of the proposed scheme, we apply the ECC-based state encoding to Instantaneous Quantum Polynomial (IQP) sampling circuits~\cite{sh09, br17, lu17}.
IQP circuits are non-Clifford and are commonly used to benchmark error-mitigation schemes on generic quantum computations.
Performance is evaluated using the normalized Cross-Entropy Benchmarking (XEB) fidelity, i.e., the raw XEB rescaled by the ideal circuit value~\cite{boi18,ar19,zl23}.
The results below show how the fidelity gain ($\Delta f$) scales with circuit size and depth.

\paragraph{Performance scaling with circuit size and depth}
\begin{figure}[t!]
    \centering
    \includegraphics[width=0.9\linewidth]{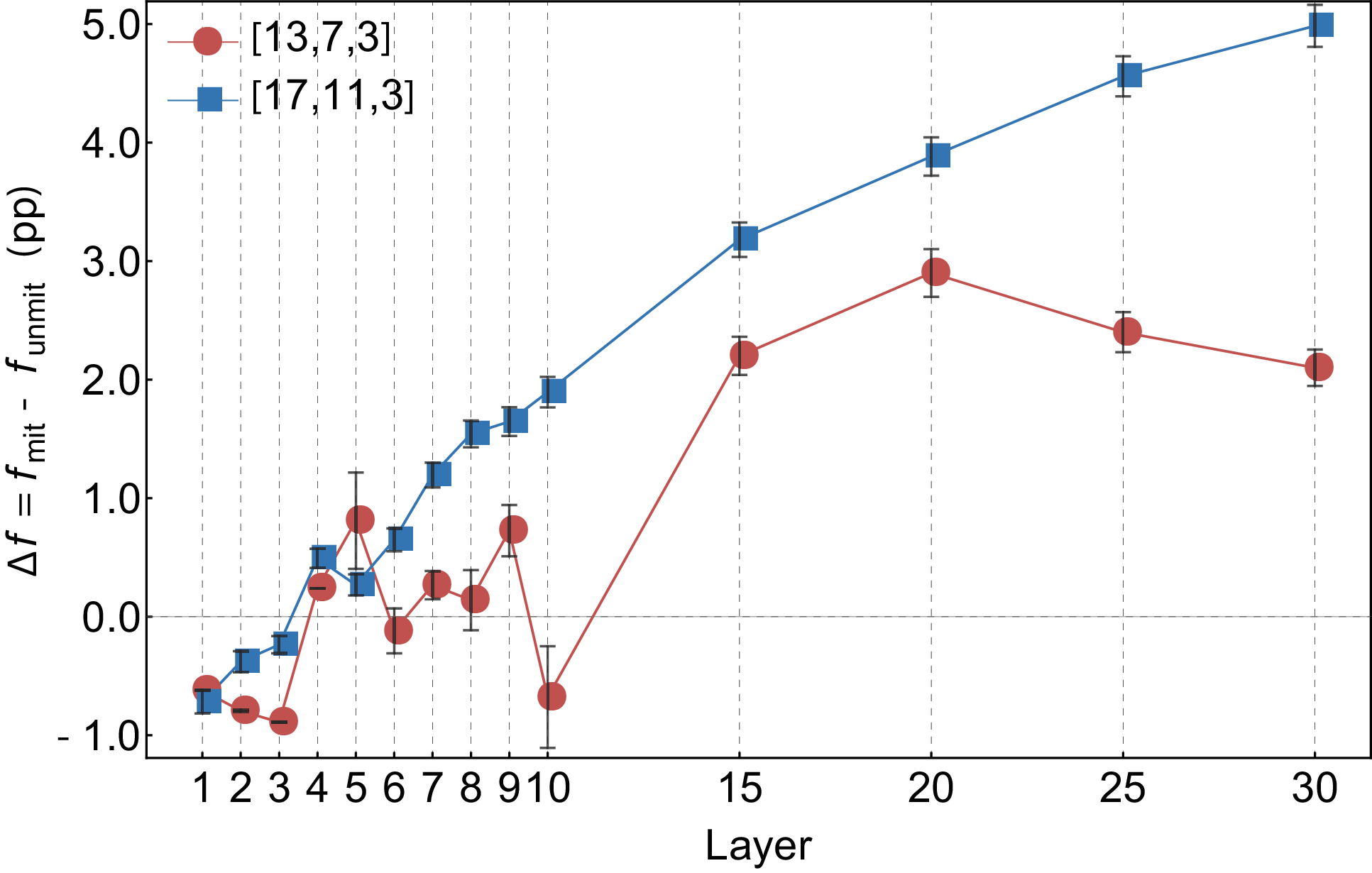}
    \caption{Fidelity gain ($\Delta f = f_{\mathrm{mit}} - f_{\mathrm{unmit}}$) of the proposed scheme on IQP sampling circuits versus circuit depth (layers) at a two-qubit gate error rate of $0.04\%$.
    Results are aggregated over 5 random seeds with 200{,}000 shots per seed (1{,}000{,}000 shots per configuration), and error bars represent one standard deviation over the seeds.
    A 7-qubit computational space $[13,7,3]$ is compared with an 11-qubit space $[17,11,3]$.
    The $[17,11,3]$ code yields larger and more stable gains.
    This improved scalability arises because the scheme adds a fixed two-qubit gate overhead (+7 for $[13,7,3]$ and +9 for $[17,11,3]$), the relative impact of which diminishes on average as circuit depth increases.}
    \label{fig:iqp_n}
\end{figure}
The efficacy of the proposed scheme reflects a trade-off between error-protection strength and the overhead from encoding and decoding, which becomes critical as size and depth grow.
Fig.~\ref{fig:iqp_n} reports the fidelity gain $\Delta f$, computed from the normalized XEB fidelity, as a function of circuit depth at a fixed two-qubit gate error rate of 0.04\%.
The 11-qubit configuration $[17,11,3]$ shows a stable improvement over the 7-qubit configuration $[13,7,3]$.
The $[13,7,3]$ code exhibits fluctuating gains and occasional loss at shallow depths, whereas the $[17,11,3]$ code maintains positive gains that increase with depth.
Consistent with these trends, Table~\ref{tab:iqp_d} shows that the proposed scheme adds a fixed two-qubit gate overhead of $+7$ for the $[13,7,3]$ code and $+9$ for the $[17,11,3]$ code.
Because this overhead is fixed, its relative impact amortizes with depth, decreasing from $63.6\%$ to $3.3\%$ for $[13,7,3]$ and from $47.4\%$ to $2.7\%$ for $[17,11,3]$ between 1 and 30 layers.
This amortization reduces the impact of encoding and decoding on the larger configuration and helps explain the more stable, depth-robust gains observed for $[17,11,3]$.
\begin{table}[h!]
    \centering
    \captionsetup{width=0.95\textwidth}
    \caption{Two-qubit gate depth $G_{2q}$ for IQP sampling circuits by code and layer.
    Entries report the total two-qubit gate count per circuit at selected layers $L\in\{1,5,10,20,30\}$ under all-to-all connectivity.
    Here, “Unmit.” denotes the baseline circuit and “Mit.” denotes the circuit encoded by the proposed scheme.
    Codes $[13,7,3]$ $(k=7)$ and $[17,11,3]$ $(k=11)$ specify the computational space size.
    The mitigated circuits add a fixed overhead relative to the unmitigated case ($+7$ for $[13,7,3]$, $+9$ for $[17,11,3]$ across the shown depths).}
    \label{tab:iqp_d}
    \begin{tabular}{l l c c c c c}
        \toprule
        \multirow{2}{*}{\vspace*{-1.6ex}Code} & \multirow{2}{*}{\vspace*{-1.6ex}\makecell{Mitigation \\ Status}}
        & \multicolumn{5}{c}{2Q gate depth at layer} \\
        \cmidrule(lr){3-7}
        & & 1 & 5 & 10 & 20 & 30 \\
        \midrule
        \multirow{2}{*}{$[13,7,3]$} & Unmit. & 11 & 39 & 74 & 144 & 214 \\
                                  & Mit.   & 18 & 46 & 81 & 151 & 221 \\
        \midrule
        \multirow{2}{*}{[17,11,3]} & Unmit. & 19 & 63 & 118 & 228 & 338 \\
                                   & Mit.   & 28 & 72 & 127 & 237 & 347 \\
        \bottomrule
    \end{tabular}
\end{table}

\paragraph{Performance under Varying Noise Levels for the \texorpdfstring{$[17,11,3]$}{[17,11,3]} Code}
\begin{figure}[h!]
    \centering
    \begin{subfigure}[b]{0.88\textwidth}
        \centering
        \includegraphics[width=\textwidth]{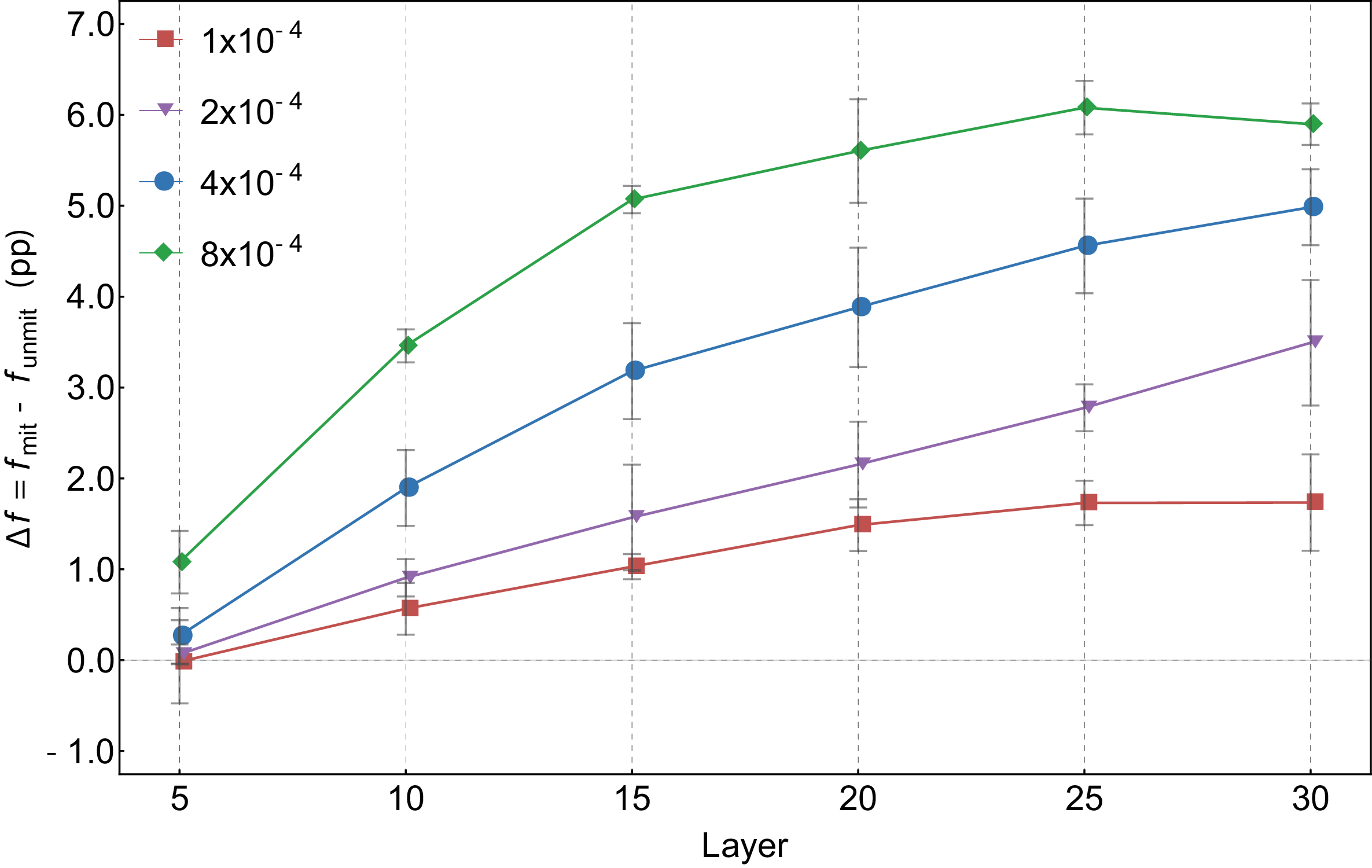}
        \caption{Fidelity gain ($\Delta f$) vs. layer depth.}
        \label{fig:iqp_gev}
    \end{subfigure}

    \begin{subfigure}[b]{0.88\textwidth}
        \centering
        \includegraphics[width=\textwidth]{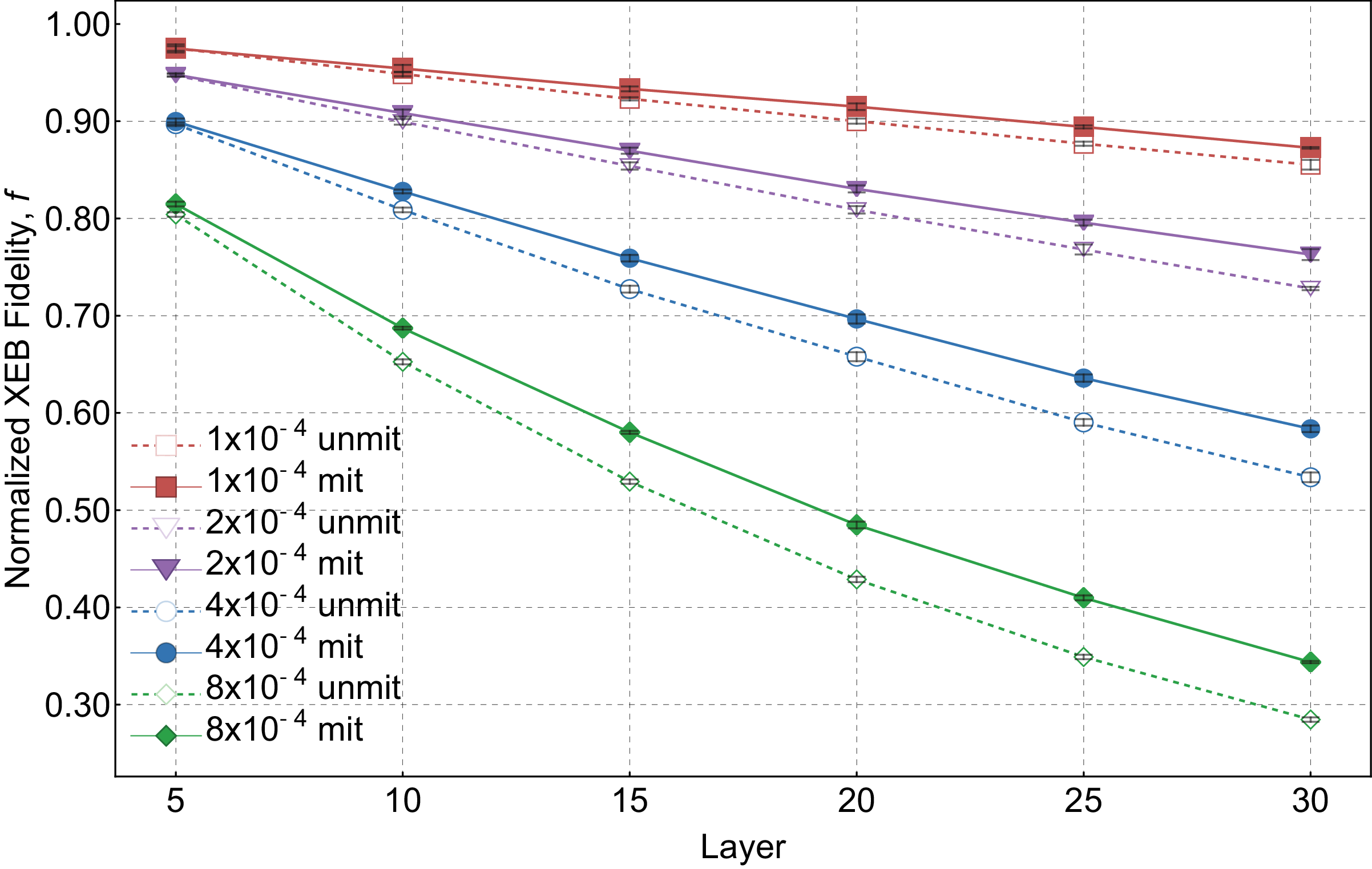}
        \caption{Normalized XEB fidelity ($f$) vs. layer depth.}
        \label{fig:abs_f}
    \end{subfigure}
    \caption{Performance of the proposed scheme on the $[17, 11, 3]$ IQP circuit across varying error rates.
    Each data point is an average over five seeds (200k shots each), with error bars representing one standard deviation over the seeds.
    (a) Fidelity gain ($\Delta f$) increases with both circuit depth and error rate.
    (b) Absolute fidelities show a consistent improvement for mitigated (solid) over unmitigated (dashed) circuits.}
    \label{fig:iqp_gevf}
\end{figure}
To assess the proposed scheme's effectiveness under different hardware conditions, the $[17,11,3]$ IQP circuit was simulated across a range of two-qubit gate error rates ($p_{2q}$): $1 \times 10^{-4}$, $2 \times 10^{-4}$, $4 \times 10^{-4}$, and $8 \times 10^{-4}$.
Each data point in Fig.~\ref{fig:iqp_gev} reports the mean fidelity gain ($\Delta f$) over five random seeds, with error bars representing $\pm 1\sigma$.

The results show that for all tested error rates, $\Delta f$ increases with circuit depth ($L$), as detailed in Fig.~\ref{fig:iqp_gev}.
The proposed scheme's advantage becomes evident at a moderate depth of $L=10$, where the fidelity gains were +0.91, +1.89, and +3.46 pp for $p_{2q} = 2\times10^{-4}, 4\times10^{-4}$, and $8\times10^{-4}$, respectively.
These gains grew substantially to +2.78, +4.56, and +6.08 pp near the end of the simulation at $L=25$.
As shown in Fig.~\ref{fig:abs_f}, this consistent positive gain translates into a clear separation in absolute fidelity ($f$) between the mitigated (solid lines) and unmitigated (dashed lines) circuits.

A key observation is the relationship between the error rate and the trend of the fidelity gain.
The curve for the highest error rate ($8 \times 10^{-4}$) exhibits a steep initial rise followed by mild saturation.
This pattern suggests that the proposed scheme aggressively corrects frequent error patterns in high-noise environments, yielding large absolute gains that level off as residual errors accumulate with depth.
Conversely, at lower error rates, the unmitigated baseline fidelity is higher, resulting in a smaller absolute gain, but the increase of $\Delta f$ with respect to $L$ is more linear.
These observations are consistent with the benchmark results of Grover's search algorithm in Fig.~\ref{fig:perf_n}.
This behavior highlights the practical utility of the proposed scheme, confirming it as a viable approach to enhancing algorithmic fidelity on hardware platforms where performance is limited by noise.

\section*{Discussion}
In this work, we introduced a simple yet structural error mitigation protocol that applies classical ECCs to the input states of quantum algorithms.
By embedding redundancy through systematic codewords, the proposed scheme alters the quantum state space to improve noise robustness, while also avoiding the need for quantum decoding, mid-circuit measurements, or syndrome extraction.

Our simulations confirm that this encoding reshapes the measurement distribution in a way that enhances the likelihood of correct outcomes.
In particular, the output probability is biased toward valid codewords, enabling reliable classical decoding even under realistic noise conditions. This per-shot correctability, without relying on aggregate distributions, distinguishes our protocol from existing statistical error mitigation methods and makes it well-suited for scenarios with sparse sampling or real-time feedback constraints.
Furthermore, our analysis revealed a critical trade-off between the code's error-correction capability and its implementation overhead.
We showed that a moderately strong code with a lower gate count can outperform a stronger code with higher overhead, highlighting that the optimal choice of code is noise-dependent and requires careful consideration of the hardware's characteristics.

The protocol is especially effective for quantum algorithms defined over computational basis states, such as Grover-like amplitude amplification and certain combinatorial optimization routines.
Beyond these specific applications, its role as a low-overhead supplement to QEC is particularly relevant in the early fault-tolerant regime.
When applied to computations on logical qubits, the scheme can process residual logical errors manifesting as bit-flips via classical post-processing of measurement outcomes.
This approach does not interfere with the resource-intensive QEC cycles, underscoring that QEC provides the primary protection during computation while the proposed protocol refines the final results.

Looking ahead, we highlight several promising directions for future research.
One avenue is to explore more expressive code families, such as nonlinear or non-systematic codes, to further optimize the trade-offs among redundancy, circuit depth, and decodability.
Another highly promising direction is to integrate the proposed protocol with modern iterative decoding techniques.
While this work utilizes hard-decision decoding, performance could be enhanced by using soft-information, for instance by employing classical Low-Density Parity-Check (LDPC) codes with iterative decoders.
Furthermore, adapting this protocol to hardware with limited connectivity and investigating its application as a supplementary processing layer for logical qubit measurements in early fault-tolerant architectures remain valuable avenues for investigation.

More broadly, our findings demonstrate that classical coding principles can be reinterpreted as quantum-structural tools, extending their role beyond post-processing to the coherent domain of algorithmic execution.
This opens a new paradigm in error mitigation: embedding classical redundancy into quantum computational fabric.
We expect that this design principle will inform practical architectures and methodologies for computing in the early fault-tolerance regime.

\section*{Methods}
\subsection*{Simulator, connectivity, and noise model}
\label{sys}
All simulations assume all-to-all connectivity and use the Qiskit Aer simulator~\cite{ja24}.
Circuits are compiled to a gate family reflecting the native set of IonQ Forte~\cite{ch24, IQNG}, with simulator-defined noise parameters anchored to its typical error metrics: $\{\mathrm{r_x},\mathrm{r_z},\mathrm{r_y},\mathrm{r_{zz}},\mathrm{id}\}$.
The noise model is based on the typical error rates reported for the IonQ Forte system (single-qubit $\sim 0.02\%$, two-qubit $\sim 0.4\%$, readout $\sim 0.5\%$)~\cite{ch24, IQFPage}.
A preliminary benchmark, presented in the~\nameref{res} section, uses these reference rates directly.

The main analysis is conducted across four lower noise levels, corresponding to the two-qubit gate error rates $p_{2q} \in \{0.08\%, 0.04\%, 0.02\%, 0.01\%\}$.
For each level, the single-qubit ($p_{1q}$) and readout ($p_{ro}$) error probabilities are scaled proportionally, as shown in equation~\ref{err_prop}.
This approach allows us to project the scheme's performance on future, higher-fidelity hardware.
It relies on the physically-motivated assumption that the relative error rates between different gate types will remain similar to those of current ion-trap systems.

Independent depolarizing channels are applied after each one-qubit ($\mathrm{r_x}$, $\mathrm{r_y}$) and two-qubit ($\mathrm{r_{zz}}$) gate.
Consistent with frame-update (“virtual-$Z$”) compilation, single-qubit $Z$ rotations are realized as software phase advances with no additional physical pulse, and we therefore do not attach a separate single-qubit depolarizing channel to $r_z$~\cite{mc17}.
A symmetric bit-flip error is also applied to each qubit upon readout.
The proportional scaling between error probabilities is defined as:
\begin{equation}
\label{err_prop}
p_{1q} = 0.05 \times p_{2q} \qquad \text{and} \qquad p_{\mathrm{ro}} = 1.25 \times p_{2q}.
\end{equation}

\subsection*{Encoding and operator implementation}
We summarize only the operational rules used in the experiments; see the Protocol description in the Results section and Fig.~\ref{fig:proex} for a detailed derivation and circuit view.
\begin{equation}
\begin{aligned}
|C(x)\rangle &= |x\rangle \otimes |p(x)\rangle,\\
O_D^E &= O_D \otimes I_{n-k}, \qquad
O_{ND}^E = U_E\, (O_{ND}\otimes I_{n-k}) \, U_E^\dagger.
\end{aligned}
\end{equation}
All measurement outcomes are classically decoded per shot as specified in the Decoding subsection.

\subsection*{Decoding and Acceptance}
\begin{algorithm}[t!]
\caption{Syndrome based decoding}
\label{alg:syn}
\KwIn{$H\in\{0,1\}^{m\times n}$ (parity-check), correction capability $t=\lfloor(d-1)/2\rfloor$, data-extraction map \textsc{ExtractString}$(\cdot)$.}
\KwOut{On each shot, corrected $k$-bit string if accepted, else \textsc{reject}.}
\BlankLine
\tcc{Offline: build a syndrome table.}
$\mathrm{ST}\leftarrow\{\mathbf 0_m \mapsto [\,]\}$.
\For{$w \leftarrow 1$ \KwTo $t$}{
  \For{each index set $I\subseteq\{0,\dots,n{-}1\}$ with $|I|=w$}{
    $e\leftarrow \mathbf 0_n$; set $e[i]\leftarrow 1$ for all $i\in I$.
    $\sigma \leftarrow (H\cdot e)\bmod 2$.
    \If{$\sigma\notin \mathrm{ST}$}{
      $\mathrm{ST}[\sigma]\leftarrow \mathrm{list}(I)$. \tcp*{store first-seen minimal-weight pattern}
    }
  }
}
\textbf{(Uniqueness extension).} Initialize multimap $\mathrm{Ext}\leftarrow \varnothing$.
\For{each index set $I\subseteq\{0,\dots,n{-}1\}$ with $|I|=t{+}1$}{
  $e\leftarrow \mathbf 0_n$; set $e[i]\leftarrow 1$ for all $i\in I$.
  $\sigma \leftarrow (H\cdot e)\bmod 2$.
  \If{$\sigma\notin \mathrm{ST}$}{append $\mathrm{list}(I)$ to $\mathrm{Ext}[\sigma]$.}
}
\For{each $\sigma$ in keys$(\mathrm{Ext})$}{
  \If{$|\mathrm{Ext}[\sigma]|=1$}{
    $\mathrm{ST}[\sigma]\leftarrow \mathrm{Ext}[\sigma][0]$. \tcp*{adopt unique $(t{+}1)$-weight pattern}
  }
}
\BlankLine
\tcc{Online: per-shot decoding with Acceptance.}
\KwIn{Measured string $v\in\{0,1\}^n$.}
$\sigma \leftarrow (H\cdot v)\bmod 2$.
\If{$\sigma\in \mathrm{ST}$}{
  $I \leftarrow \mathrm{ST}[\sigma]$.
  \For{each $i\in I$}{$v[i]\leftarrow v[i]\oplus 1$.}
  $\mathbf s \leftarrow \textsc{ExtractString}(v)$.
  \Return $\mathbf s$.
}
\Else{
  \Return \textsc{reject}.
}
\end{algorithm}
We use syndrome-based decoding with a precomputed syndrome table ($\mathrm{ST}$) of correctable patterns.
First, we enumerate all error patterns up to the guaranteed capability $t=\lfloor(d-1)/2\rfloor$~\cite{li83} and record their syndromes in the $\mathrm{ST}$, keeping the first-seen minimal-weight pattern for each syndrome.
We then extend $\mathrm{ST}$ with weight-$(t{+}1)$ patterns only when the corresponding syndrome appears uniquely among all such candidates, which avoids degeneracy while enabling corrections beyond the nominal distance.
At run time, each measured $n$-bit string $v$ is accepted if its syndrome $\sigma=Hv\bmod 2$ is present in $\mathrm{ST}$, in which case we flip the listed positions and extract the $k$-bit string; otherwise the shot is rejected.
The detailed procedure for this decoding and Acceptance process is outlined in Algorithm~\ref{alg:syn}.

\subsection*{Circuit families}
\paragraph*{Grover's search algorithm.}
We use an amplitude-amplification structure with a phase-flip oracle and a diffusion operator~\cite{gr96, gr97}. In the encoded version, the Oracle is diagonal and applied as $O_D^E$, while the diffusion is implemented as $O_{ND}^E$ via $U_E$ and $U_E^\dagger$.
The number of Grover iterations, $R$, is set to the optimal value for a single marked item, given by $R = \lfloor\frac{\pi}{4}\sqrt{2^k}\rfloor$, where $k$ is the number of computational qubits.
Unless stated otherwise, Grover experiments use $4{,}000$ shots per circuit and are averaged over 10 random seeds, matching the configurations reported in the Results figures and tables.

\paragraph*{IQP sampling.}
We instantiate IQP circuits~\cite{sh09, br17, lu17} with a single global Hadamard layer at the beginning, a stack of $L$ identical diagonal layers $D_Z$, and a final global Hadamard layer under all-to-all connectivity.
\begin{equation}
U_{\mathrm{IQP}}(L) \;=\; H^{\otimes k}\, D_Z^{\,L}\, H^{\otimes k}.
\end{equation}
Each diagonal layer is fixed and reused across the stack,
\begin{equation}
D_Z \;=\; \Big(\prod_{i=1}^{k} R_Z(\theta_z^{(i)})\Big)\,\Big(\prod_{1\le i<j\le k} R_{ZZ}^{(i,j)}(\theta_{zz}^{(i,j)})\Big),
\end{equation}
where the pair set covers all unordered qubit pairs.
The angle vectors $\boldsymbol{\theta}_z$ and $\boldsymbol{\theta}_{zz}$ are predefined once (hard-coded) and shared between the unmitigated and mitigated circuits, yielding a time-invariant diagonal block composed $L$ times.
For implementation, we prebind $(\boldsymbol{\theta}_z,\boldsymbol{\theta}_{zz})$ to a template diagonal layer and compose this bound layer $L$ times between the two Hadamard layers.
In the mitigated circuit, the diagonal blocks are applied directly to the computational qubits as $D_Z\otimes I_{n-k}$, while each global Hadamard is applied via conjugation as $U_E\,(H^{\otimes k}\!\otimes I_{n-k})\,U_E^\dagger$.
All IQP experiments use $200{,}000$ shots per configuration and are repeated five times with independent seeds, and we report per-run values and their averages.
Furthermore, unmitigated and mitigated instances form matched pairs by sharing the same angle vectors and compilation settings.
Performance is evaluated using the normalized Cross-Entropy Benchmarking (XEB) fidelity $f$~\cite{boi18,ar19,zl23}, and we summarize the improvement as $\Delta f = f_{\mathrm{mit}}-f_{\mathrm{unmit}}$.

\subsection*{Metrics}
\paragraph*{Acceptance.}
As defined in the Decoding and Acceptance subsection, a shot is accepted iff its syndrome $\sigma = H v \bmod 2$ is present in the precomputed table $\mathrm{ST}$.
We denote by $N_{\mathrm{total}}$ the total number of executed measurement shots before any post-selection.
We denote by $N_{\mathrm{accepted}}$ the number of shots retained after post-selection, i.e., those whose syndromes are present in $\mathrm{ST}$ and are therefore kept for decoding.
For completeness, the number of rejected shots is $N_{\mathrm{rejected}} = N_{\mathrm{total}} - N_{\mathrm{accepted}}$.
The Acceptance is defined as
\begin{equation}
A = \frac{N_{\mathrm{accepted}}}{N_{\mathrm{total}}} \times 100
\end{equation}

\paragraph*{Success probability (Grover's search algorithm).}
The success probability is the fraction of shots yielding the target string after decoding.
We distinguish the conditional and unconditional success probabilities.
They are defined as
\begin{equation}
P_{s|acc} \;=\; \frac{N_{\mathrm{correct}}}{N_{\mathrm{accepted}}},
\qquad
P_{s} \;=\; \frac{N_{\mathrm{correct}}}{N_{\mathrm{total}}},
\end{equation}
where $N_{correct}$ is the number of shots yielding the target string after decoding.
For the proposed scheme, the end-to-end relation is
\begin{equation}
P_{s}^{\mathrm{mit}} \;=\; \frac{A}{100} \cdot P_{s|acc}^{\mathrm{mit}}.
\end{equation}
For the baseline, all shots are accepted.
Hence
\begin{equation}
A^{\mathrm{base}} = 100,
\qquad
P_{s|acc}^{\mathrm{base}} = P_{s}^{\mathrm{base}}.
\end{equation}

\paragraph*{Normalized XEB (IQP).}
Let $p_{\mathrm{ideal}}(x)$ be the ideal distribution.
The raw XEB is
\begin{equation}
\mathrm{XEB} \;=\; 2^k\,\mathbb{E}_{x\sim p_{\mathrm{exp}}}\!\left[p_{\mathrm{ideal}}(x)\right] - 1.
\end{equation}
We report the normalized fidelity, $f$, and the fidelity gain, $\Delta f$, which is the difference between the fidelity of the mitigated circuit ($f_{\mathrm{mit}}$) and the unmitigated circuit ($f_{\mathrm{unmit}}$)~\cite{boi18,ar19,zl23}.
These are defined as:
\begin{equation}
f \;=\; \frac{\mathrm{XEB}}{S_{\mathrm{ideal}}},\qquad
S_{\mathrm{ideal}} \;=\; 2^k\,\mathbb{E}_{x\sim p_{\mathrm{ideal}}}\!\left[p_{\mathrm{ideal}}(x)\right] - 1,\qquad
\Delta f \;=\; f_{\mathrm{mit}} - f_{\mathrm{unmit}}.
\end{equation}

\subsection*{Parity-check matrices used in simulations and encoding operator}
All matrices are over the finite field $\mathbb{F}_2$.
We display $H$ in systematic form $H=[P^\top\,|\,I_{n-k}]$~\cite{li83}.
State preparation initializes the $(n-k)$ parity qubits to $|0\rangle$ and the $k$ data qubits to the computational register.
Then the encoder $U_E$ implements the map $x \mapsto [x\,|\,xP]$ over $\mathbb{F}_2$ by applying $\mathrm{CNOT}$ from data qubit $i$ to parity qubit $j$ for each $P_{i,j}=1$.
\begingroup
\small
\setlength{\arraycolsep}{4pt}

\paragraph*{$[11,7,3]$ code.}
\begin{equation}
H_{11,7,3} \;=\;
\left[
\begin{array}{ccccccc|cccc}
0&0&0&0&1&1&1&1&0&0&0\\
0&1&1&1&0&0&1&0&1&0&0\\
1&0&1&1&0&1&0&0&0&1&0\\
1&1&0&1&1&0&0&0&0&0&1
\end{array}
\right].
\end{equation}

\paragraph*{$[13,7,3]$ code.}
\begin{equation}
H_{13,7,3} \;=\;
\left[
\begin{array}{ccccccc|cccccc}
1&0&0&0&1&0&0& 1&0&0&0&0&0\\
0&1&0&0&0&1&0& 0&1&0&0&0&0\\
0&0&1&0&0&0&1& 0&0&1&0&0&0\\
0&0&0&1&1&0&1& 0&0&0&1&0&0\\
1&0&1&0&0&1&0& 0&0&0&0&1&0\\
0&1&0&1&0&0&0& 0&0&0&0&0&1
\end{array}
\right].
\end{equation}

\paragraph*{$[15,7,3]$ code.}
\begin{equation}
H_{15,7,3} \;=\;
\left[
\begin{array}{ccccccc|cccccccc}
1&0&0&0&1&0&0& 1&0&0&0&0&0&0&0\\
0&1&0&0&0&1&0& 0&1&0&0&0&0&0&0\\
0&0&1&0&0&0&1& 0&0&1&0&0&0&0&0\\
0&0&0&1&0&0&0& 0&0&0&1&0&0&0&0\\
1&0&0&0&0&0&0& 0&0&0&0&1&0&0&0\\
0&1&0&0&1&0&0& 0&0&0&0&0&1&0&0\\
0&0&1&0&0&1&0& 0&0&0&0&0&0&1&0\\
0&0&0&1&0&0&1& 0&0&0&0&0&0&0&1
\end{array}
\right].
\end{equation}

\paragraph*{$[15,7,5]$ code.}
\begin{equation}
H_{15,7,5} \;=\;
\left[
\begin{array}{ccccccc|cccccccc}
1&1&0&1&0&0&0&1&0&0&0&0&0&0&0\\
0&1&1&0&1&0&0&0&1&0&0&0&0&0&0\\
0&0&1&1&0&1&0&0&0&1&0&0&0&0&0\\
0&0&0&1&1&0&1&0&0&0&1&0&0&0&0\\
1&1&0&1&1&1&0&0&0&0&0&1&0&0&0\\
0&1&1&0&1&1&1&0&0&0&0&0&1&0&0\\
1&1&1&0&0&1&1&0&0&0&0&0&0&1&0\\
1&0&1&0&0&0&1&0&0&0&0&0&0&0&1
\end{array}
\right].
\end{equation}

\paragraph*{$[17,7,3]$ code.}
\begin{equation}
H_{17,7,3} \;=\;
\left[
\begin{array}{ccccccc|cccccccccc}
1&0&0&0&0&1&0&1&0&0&0&0&0&0&0&0&0\\
0&1&0&0&0&0&0&0&1&0&0&0&0&0&0&0&0\\
0&0&1&0&0&0&1&0&0&1&0&0&0&0&0&0&0\\
1&0&0&0&0&0&0&0&0&0&1&0&0&0&0&0&0\\
0&1&0&0&0&0&0&0&0&0&0&1&0&0&0&0&0\\
0&0&1&0&0&0&0&0&0&0&0&0&1&0&0&0&0\\
0&0&0&1&0&1&0&0&0&0&0&0&0&1&0&0&0\\
0&0&0&1&0&0&0&0&0&0&0&0&0&0&1&0&0\\
0&0&0&0&1&0&1&0&0&0&0&0&0&0&0&1&0\\
0&0&0&0&1&0&0&0&0&0&0&0&0&0&0&0&1
\end{array}
\right].
\end{equation}

\paragraph*{$[17,11,3]$ code.}
\begin{equation}
H_{17,11,3} \;=\;
\left[
\begin{array}{ccccccccccc|cccccc}
1&0&0&1&0&1&0&1&0&0&0& 1&0&0&0&0&0\\
1&0&0&0&1&0&1&0&1&0&0& 0&1&0&0&0&0\\
0&1&0&1&0&0&1&0&0&1&0& 0&0&1&0&0&0\\
0&1&0&0&1&1&0&0&0&0&1& 0&0&0&1&0&0\\
0&0&1&0&0&0&0&1&0&1&0& 0&0&0&0&1&0\\
0&0&1&0&0&0&0&0&1&0&1& 0&0&0&0&0&1
\end{array}
\right].
\end{equation}
\endgroup

\subsection*{Statistics and reproducibility}
All simulated data points are presented as the mean over a number of independent random seeds, with error bars in all figures representing one standard deviation over these seeds.
For the Grover's search algorithm experiments, results were aggregated over 10 independent random seeds, with 4,000 measurement shots per seed.
For the IQP sampling circuit experiments, results were aggregated over 5 independent random seeds, with 200,000 measurement shots per seed.
No formal statistical significance tests were performed; instead, performance differences are reported as absolute gains in percentage points (pp).

\section*{Data availability}
All relevant data and figures supporting the main conclusions of the document are available from the corresponding author upon request.

\section*{Code availability}
All relevant code supporting the document is available from the corresponding author upon request.

\subsection*{Acknowledgements}
This research was supported by Korea Institute of Science and Technology Information(KISTI).(No.(KISTI)K25L5M2C2, (NTIS)2710087116), This research was supported by QuantumComputing based on QuantumAdvantage challenge research through the National Research Foundation of Korea (NRF) funded by the Korean government(MSIT) (RS-2023-00256221).

\subsection*{Author contributions}
I.S. conceived the study, developed the core methodology, analyzed the results, and wrote the manuscript.
C.L., W.S., and K.B. performed the simulations and contributed to the refinement of the scheme.
W.L. provided supervision and guidance during the work.
All authors contributed to the discussions of the results and reviewed the final manuscript.

\subsection*{Competing Interests}
The authors declare no competing interests.

\end{document}